\documentclass[11pt,twoside]{article}
\usepackage{asp2014}

\aspSuppressVolSlug
\resetcounters

\begin{document}

\title{Checkpoint, Restore, and Live Migration for Science Platforms}

\author{Mario~Juric, Steven~Stetzler, and Colin~T.~Slater}
\affil{DiRAC Institute and the Department of Astronomy, University of Washington, Seattle, WA, U.S.A; \email{mjuric@astro.washington.edu}}

\paperauthor{Mario~Juric}{mjuric@astro.washington.edu}{0000-0003-1996-9252}{University of Washington}{DiRAC Institute and the Department of Astronomy}{Seattle}{WA}{98195}{U.S.A.}
\paperauthor{Steven~Stetzler}{stevengs@uw.edu}{0000-0002-7712-6678}{University of Washington}{DiRAC Institute and the Department of Astronomy}{Seattle}{WA}{98195}{U.S.A.}
\paperauthor{Colin~T.~Slater}{ctslater@uw.edu}{0000-0002-0558-0521}{University of Washington}{DiRAC Institute and the Department of Astronomy}{Seattle}{WA}{98195}{U.S.A.}



  
\begin{abstract}

We demonstrate a fully functional implementation of (per-user) checkpoint, restore, and live migration capabilities for JupyterHub platforms. Checkpointing -- the ability to freeze and suspend to disk the running state (contents of memory, registers, open files, etc.) of a set of processes -- enables the system to snapshot a user's Jupyter session to permanent storage. The restore functionality brings a checkpointed session back to a running state, to continue where it left off at a later time and potentially on a different machine. Finally, live migration enables moving running Jupyter notebook servers between different machines, transparent to the analysis code and w/o disconnecting the user. Our implementation of these capabilities works at the system level, with few limitations, and typical checkpoint/restore times of O(10s) with a pathway to O(1s) live migrations. It opens a myriad of interesting use cases, especially for cloud-based deployments: from checkpointing idle sessions w/o interruption of the user's work (achieving cost reductions of 4x or more), execution on spot instances w. transparent migration on eviction (with additional cost reductions up to 3x), to automated migration of workloads to ideally suited instances (e.g. moving an analysis to a machine with more or less RAM or cores based on observed resource utilization). The capabilities we demonstrate can make science platforms fully elastic while retaining excellent user experience.
  
\end{abstract}

\section{Introduction}

With ever-increasing dataset sizes, remote analysis paradigms are becoming increasingly popular. In such systems \citep[e.g.][]{LSP,2020A&C....3300412T,2020A&C....3300411N,StetzlerGateways}, the users access data and computing resources through \textit{science platforms} -- rich gateways exposing server-side code editing, management, execution and result visualization capabilities -- usually implemented as {\em notebooks} such as Jupyter \citep{jupyter}, or Zeppelin. A challenge of this model is that the data provider (e.g., an archive facility) now bears both the cost of dataset storage and that of computing resources -- including those used for running the users' Jupyter notebooks. This cost can balloon quickly, especially on cloud resources: left unmanaged, a 24/7 run for a 100 users reaches \$$300,000+$ range. This can be reduced by  terminating inactive instances, but the price is a poor user experience.


In this contribution we present a solution: the ability to checkpoint (freeze) a user's running Jupyter notebook server to disk, and restore it to memory on-demand (including on a different host). This {\em C/R} functionality can dramatically reduce the cost, while fully maintaining the user experience. It also enables novel capabilities, such as uninterrupted migration of work based on resource needs.

\section{Elsa: A Checkpoint-able JupyterHub Deployment}

In a fully functional proof-of-concept we named {\em Elsa}\footnote{The source code is available at \url{https://github.com/dirac-institute/elsa}}, we added the C/R functionality to a cloud deployment of JupyterHub. The user experience can be viewed in a YouTube screencast at \url{https://dirac.us/5aj}; here, we provide a brief summary.
\\


\begin{wrapfigure}{l}{0.5\textwidth}
	\begin{center}
		\vspace{-10pt}
		\includegraphics[width=0.48\textwidth]{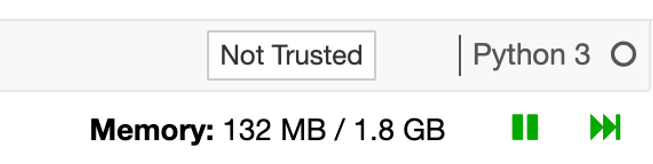}
		\vspace{10pt}
		\caption{C/R UI. Pause checkpoints, fast-forward initiates migration to a different VM instance type.
			\label{fig:ui}}
		\vspace{-10pt}
	\end{center}
\end{wrapfigure}

\noindent With Elsa, the user logs into the JupyterHub aspect of the science platform and starts Jupyter on a machine with desired capabilities (e.g., CPU core count, or RAM size). The user then works on their notebooks as usual. However, an additional option is now present in the notebook interface -- the ``pause'' button on the right of the notebook toolbar (Figure~\ref{fig:ui}). Clicking this button checkpoints the complete state (memory, open files, etc.)  of the notebook server, and releases all  computational resources. Later, the user can restore the checkpointed session either on the same VM or one with different resources, and continue where they left off as if nothing happened.


\section{Implementation}

\begin{figure}
	\centering
	\includegraphics[width=0.80\textwidth]{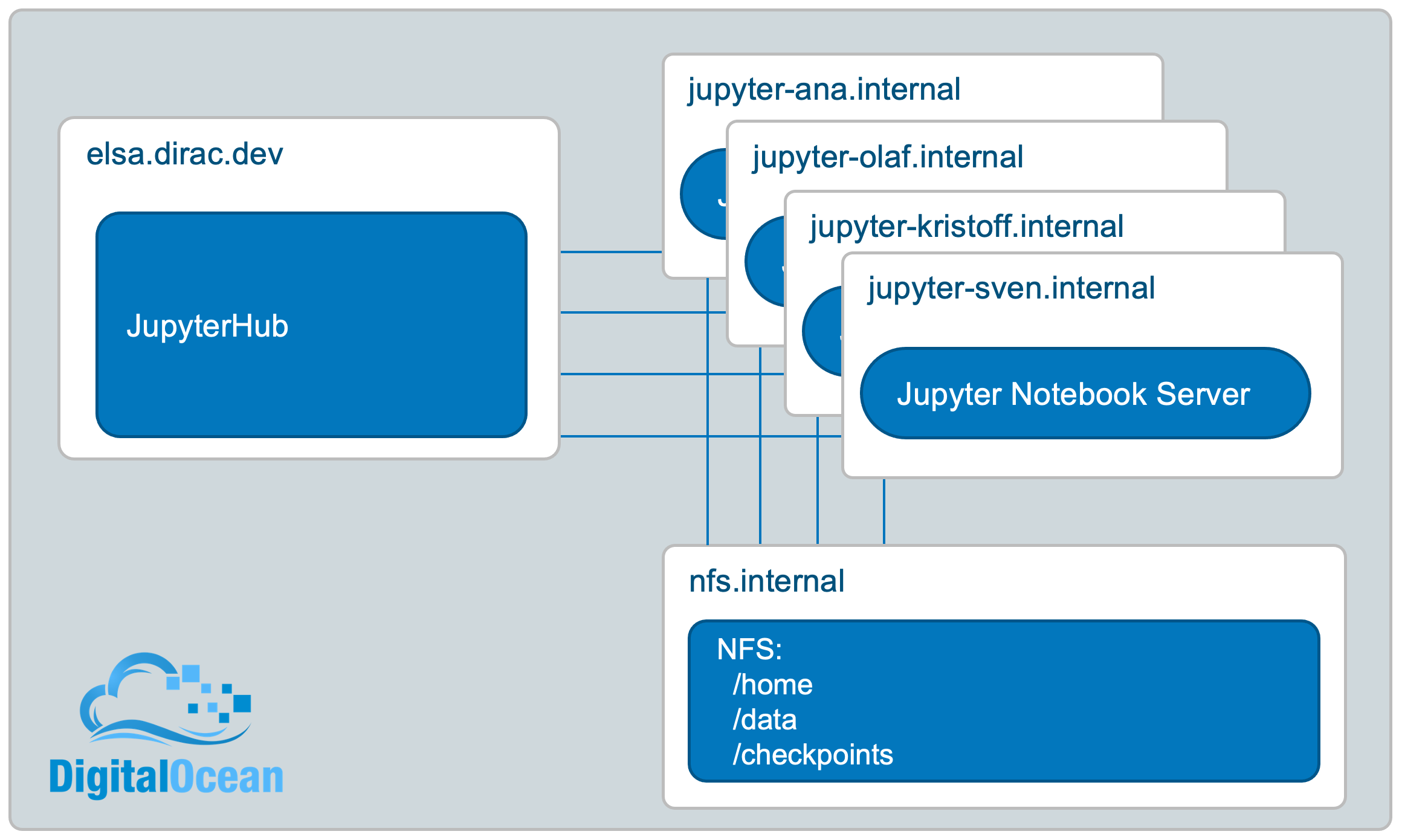}
	\caption{The architecture of the checkpointable JupyterHub deployment. \label{fig:architecture}}
\end{figure}

The high-level architecture of the Elsa prototype is shown in Figure~\ref{fig:architecture}. The  JupyterHub front-end is run on a single, dedicated, node, from an essentially unmodified upstream JupyterHub container image.

Our main (configuration-level) customization is the addition of a new {\tt Spawner} class. Within JupyterHub, a spawner is responsible for starting and managing users' notebook server instance(s)\footnote{See \url{https://jupyterhub.readthedocs.io/en/stable/reference/spawners.html} for more.}. The spawner finds a VM where the notebooks run, and starts and stops Jupyter on that VM. As it has no support for pod C/R, we could not use Kubernetes at this time\footnote{Work is ongoing; see \url{https://github.com/kubernetes/enhancements/pull/1990}}; instead, our spawner directly allocates one new VM per user from the cloud provider. While reducing portability, running on bare VMs does bring some additional (and significant) benefits\footnote{Isolation between users at the VM level leads to predictable user experience; using VMs allows us to add swap making out-of-memory conditions a ``soft'' fail; and bare VMs are faster to provision relative to speeds from common {\tt k8s} cluster autoscalers.}.

Although each user gets their own VM, per-user Jupyter is still run from a container. This i) abstracts away the details of the raw VM (e.g., Linux distribution doesn't matter, as long as podman/CRIU are available), ii) allows us to use the standard notebook-server container, iii) makes deployment significantly easier (a simple {\tt pull}, rather than OS-level install), and iv) allows for secure re-use of VMs between different users (as users are sandboxed by their container).

We manage the container using {\tt podman}\footnote{A daemonless container engine for OCI Containers; \url{https://podman.io/}}, which has built in support for container checkpointing with {\tt CRIU}\footnote{Checkpoint-and-Restore in Userspace; \url{http://criu.org}}. To inject C/R support into Jupyter, we hijack the {\tt Spawner's} start and stop APIs. When receiving the start command, our custom spawner restores the session from a previously stored checkpoint, if one exists. Similarly, upon receiving a request to stop it checkpoints rather than stops. This is clearly a convenient hack, and a proper C/R API should ultimately be added to JupyterHub.

Each VM mounts disks from a shared NSF server ({\tt nsf.internal} in Figure~\ref{fig:architecture}), including the {\tt /home} filesystem which is mounted as a volume within the user's container. A shared {\tt /home} elegantly solves the problem of how to keep users' data identical on an inode level if/when they restore a checkpoint on a different machine (a requirement for checkpointing). While a shared filesystem introduces a potential bottleneck (e.g., imagine thousands of users simultaneously analyzing large datasets from their homedirs), we haven't observed any issues in typical usage. For larger deployments, one could use a more scalable shared filesystem (e.g. {\tt pNFS} or {\tt GPFS}). We use the same NFS to centrally store the checkpoints themselves.

Finally, this new functionality is exposed to the user through a simple two-button UI featuring a ``pause'' and ``fast forward'' buttons (Figure~\ref{fig:ui}). For simplicity, we added these to the {\tt nbresuse} Jupyter extension.

This code has been prototyped and deployed on Digital Ocean\footnote{Digital Ocean is a low-cost cloud provider;  \url{https://www.digitalocean.com/}} ({\tt DO}). Given we use {\tt DO} APIs in our spawner to manage the VM instances, Elsa will not run on other providers out-of-the-box. However, extensions to other clouds (e.g., AWS or GCP) are rather easy -- O(50) lines of code -- and planned as a future addition.

\section{Discussion and Future Work}

To our knowledge, this is the first fully functional implementation of checkpoint-restore and migration functionality of Jupyter notebooks on JupyterHub. It demonstrates that C/R for Jupyter not only possible, but fully functional for analyses as complex as the LSST software stack\citep{2017ASPC..512..279J}. Going forward we plan to generalize the code to other Cloud providers, implement migration of open TCP network connections, add a dedicated C/R API for JupyterHub, and improve overall C/R performance.

We see three main application areas for this work: shared servers, on-prem science platforms, and Cloud-based science platforms. For shared servers (e.g., a machine used by a research group), one can now opportunistically checkpoint Jupyter instances after a period of inactivity thus optimizing overall resource usage. For on-prem science platforms our work lets the platform operator checkpoint rather than terminate inactive instances resulting in significantly better user experience. They can also dynamically migrate users' instances to optimize resource usage.

But the largest opportunity is for Cloud deployments, where C/R can both improve the user experience and  significantly lower the operating cost. As we show in Table~1, with our C/R work running a typical user's Jupyter instance may cost as little as \$$200$/yr, with no degradation to user experience relative to running 24/7. This is likely better than the total cost of ownership of running a similar system on premise, eliminating one more barrier to migrating science analyses to the Cloud.

\begin{table}
	\centering
	\begin{tabular}{l r} 
		\hline
		Instance & Annual Cost \\
		\hline \hline
		24x7x365 8core / 32G RAM (m5.2xlarge, on-demand)    &    \$3364 \\
		As above + c/r, 15\% duty cycle                                                   &    \$505 \\
		As above + c/r, 15\% duty cycle, spot                                        &     \$196 \\
		Savings                                                                                                    &     $\sim17$x \\
		\hline
	\end{tabular}
	\caption{Savings when running on cloud resources (AWS pricing as of 3:20pm PST, Nov 4, 2020.). The first row shows the cost of running an on-demand instance for an entire year (giving ideal user experience). The second row shows the cost of running for 15\% of that time (a typical duty cycle we observed with our users), storing a checkpoint while the user is inactive. Finally we show the cost of running on spot instances, which is now possible as users work can be transparently migrated to a new instance if the spot VM is to be terminated.}
	\label{table:cloudcost}
\end{table}

\bibliography{D1-102}


\end{document}